\documentclass[twocolumn, letter]{emulateapj}

\usepackage{graphics,graphicx}
\usepackage{fancyheadings}
\usepackage{color}
\usepackage{natbib}

\newcommand{\chandra}{\textit{Chandra}}
\newcommand{\fermi}{\textit{Fermi}}
\newcommand{\pks}{PKS\,1718$-$649}
\def\arcsec{\hbox{$^{\prime\prime}$}}

\def\pasa{PASA}

\shorttitle{High-energy emission in \pks}
\shortauthors{Sobolewska et al.}

\begin{document}

\title{The origin of high-energy emission in the young radio source PKS\,1718$-$649}

\author{Ma\l{}gosia Sobolewska$^1$, Giulia Migliori$^2$, Luisa Ostorero$^{3, 4}$,
Aneta Siemiginowska$^1$, {\L}ukasz Stawarz$^5$, Matteo Guainazzi$^6$, Martin J. Hardcastle$^7$}

\affil{$^1$ Center for Astrophysics $|$ Harvard \& Smithsonian, 60 Garden Street, Cambridge, MA 02138, USA.}
\affil{$^2$ INAF, Istituto di Radio Astronomia di Bologna, Via P. Gobetti 101, I-40129 Bologna, Italy}
\affil{$^3$ Dipartimento di Fisica, Universit\`a di Torino, Via P. Giuria 1, 10125 Torino, Italy}
\affil{$^4$ Istituto Nazionale di Fisica Nucleare (INFN), Sezione di Torino, Via P. Giuria 1, 10125 Torino, Italy}
\affil{$^5$ Astronomical Observatory, Jagiellonian University, ul. Orla 171, 30-244 Krak\'ow, Poland}
\affil{$^6$ European Space Research and Technology Centre (ESA/ESTEC),
Kepleriaan 1, 2201 AZ, Noordwijk, The Netherlands}
\affil{$^7$ Department of Physics, Astronomy and Mathematics, University of Hertfordshire,
College Lane, Hatfield AL10 9AB, UK}

\smallskip
\email{msobolewska@cfa.harvard.edu}


\label{firstpage}

\begin{abstract}

We present a model for the broadband radio-to-$\gamma$-ray spectral energy distribution
of the compact radio source \pks. Because of its young age ($\simeq 100$ years) and proximity  ($z=0.014$), \pks\ offers a unique opportunity to study nuclear conditions and the jet/host galaxy feedback process at the time of the initial radio jet expansion. \pks\ is one of a handful of young radio jets with $\gamma$-ray emission confirmed with the {\it Fermi}/LAT detector. We show that this $\gamma$-ray emission can be successfully explained by Inverse Compton scattering of the ultraviolet photons, presumably from an accretion flow, off non-thermal electrons in the expanding radio lobes. The origin of the X-ray emission in \pks\ is more elusive. While Inverse Compton scattering of the infrared photons emitted by a cold gas in the vicinity of the expanding radio lobes contributes significantly to the X-ray luminosity, the data require an additional source of X-rays, e.g. a weak X-ray corona or a radiatively inefficient accretion flow, expected from a LINER type nucleus such as that of \pks. We find that the jet in \pks\ has low power, $L_j \simeq 2 \times 10^{42}$\,erg\,s$^{-1}$, and expands in an environment with density $n_0 \simeq 3 - 20$\,cm$^{-3}$. The inferred mass accretion rate and gas mass reservoir within 50-100\,pc from the galactic center are consistent with estimates from the literature obtained by tracing molecular gas in the innermost region of the host galaxy with SINFONI and ALMA.

\end{abstract}


\section{Introduction}
\label{sec:intro}

\pks\ is a well known radio source in NGC\,6328 classified as a low-ionization nuclear emission-line region
(LINER) galaxy with photoionization as the main
excitation mechanism of emission lines (Filippenko 1985). It hosts a supermassive black hole
with a mass of the order of 10$^8$\,M$_{\odot}$ (Willett et al. 2010).
The radio source has convex radio spectrum peaking
in the GHz range (Tingay et al. 2015) and it belongs to the class of
Gigahertz-Peaked Spectrum sources (GPS; e.g. O'Dea 1998; O'Dea \& Siemiginowska 2016;
O'Dea \& Saikia 2021). 
GPS sources with radio lobes that show symmetric morphology,
as in the case of \pks\ (Tingay et al. 1997), are known as Compact Symmetric Objects (CSOs;
e.g. O'Dea \& Saikia 2021; Orienti 2016; Wilkinson et al. 1994).
They appear to be
smaller versions of classical doubles (i.e., Fanaroff-Riley type II radio galaxies;
Fanaroff \& Riley 1974).
Multi-epoch radio monitoring
of the expansion of the lobes of \pks\ implies that the radio source is very young,
$t_{\rm age} \simeq 100$\,years, and small, with parsec scale linear radio size
(Polatidis \& Conway 2003; Angioni et al. 2019). At its redshift of $z=0.0144$ 
(Meyer et al. 2004), it is one of
the nearest CSOs with a measured kinematic age known to date (An \& Baan 2012).

\pks\ is currently one of the best studied examples of a newly-born radio source, observed
and detected across the whole electromagnetic spectrum, from the radio to the $\gamma$-ray band.
The source has been observed
spectroscopically in the mid infrared band (MIR) with {\it Spitzer} and showed
signatures typical of both star-forming gas and active
galactic nucleus (AGN) gas illumination
(Willett et al. 2010). Filippenko (1985) demonstrated that the optical
light of the host galaxy of \pks\ contains a contribution from a non-stellar
power-law continuum that might be associated with weak nuclear emission.
Siemiginowska et al. (2016) observed \pks\
with \chandra\ for the first time in the X-ray band and found that a point source
is embedded in extended X-ray emission that was studied in detail by
Beuchert et al. (2018). The detection of \pks\ in the $\gamma$-ray band was first
reported by Migliori et al. (2016) and then confirmed by the Fermi-LAT 4th Source catalog
(4FGL; Abdollahi et al. 2020). In general, jetted radio sources with jets pointing away
from the line-of-sight, and in particular sources symmetric in the plane of the sky,
are not expected to be strong $\gamma$-ray emitters, as opposed to blazars
in which the emission is beamed due to the jet orientation. Indeed, non-blazar
type sources constitute only $\sim$2\% of all AGNs in the 4FGL.
However, Stawarz et al.
(2008) and Ostorero et al. (2010) posited that CSO high-energy emission, in particular
$\gamma$-ray emission, is expected due to
Inverse Compton (IC) scattering of the ambient low-energy photons off the non-thermal
electron populations within the expanding radio lobes inflated by the radio jet.
While CSOs are indeed regularly detected in the X-ray band even in short exposures
(e.g. Siemiginowska et al. 2009; Sobolewska et al. 2019a; and
references therein), \pks\ remains one of only a handful of
$\gamma$-ray emitters with a firm CSO
classification to date (M{\"u}ller et al. 2014, 2016; Migliori et al. 2016;
Principe et al. 2020; Lister et al. 2020)\footnote{We note that that PMN J1603-4904 studied by
M{\"u}ller et al. (2014, 2016) is not included anymore in the latest {\it Fermi}/LAT catalog.}.

An alternative explanation of the compact radio nature of sources like
\pks\ involves a confinement by a dense inter-stellar medium (ISM) rather than a young
age (e.g. van Breugel et al. 1984; O'Dea 1998; Dicken et al. 2012). However, in this paper
we choose to explore the youth scenario motivated by the fact that \pks\ shows a 
rather low X-ray absorbing column density $N_H$ (Siemiginowska et al. 2016; Beuchert et al. 2018).
Moreover, rough estimates suggest that $10^9 - 10^{10}$\, $M{_\odot}$ gas mass would be required
to confine a radio source (O’Dea \& Saikia 2021; and references therein), while the estimated
gas mass reservoir in \pks\ is orders of magnitude lower based on both $N_H$ and H2 measurements
(O'Dea \& Saikia 2021; Maccagni et al. 2016); and multi-epoch radio observations support
a constant expansion of the radio lobes in \pks\ (Angioni et al. 2019).

We study the broadband radio-to-$\gamma$-ray emission of \pks, and
identify the physical processes that dominate the high-energy radiative
output of a radio source in formation. We first collect the multiwavelength
observations of the source to construct
its broadband spectral energy distribution (SED; Section~\ref{sec:obs}).
We then model the observed SED in the framework of the expanding radio lobe model of
Stawarz et al. (2008; Section~\ref{sec:model}).
We present our results in Section~\ref{sec:res}, we discuss our
findings in Section~\ref{sec:disc}, and conclude in Section~\ref{sec:concl}.
Throughout the paper, we use the most recent constraints on 
the cosmological parameters to convert the
observed fluxes into luminosities (Hinshaw et al. 2013; $\rm H_0 = 69.3$\,km\,s$^{-1}$\,
Mpc$^{-1}$, $\rm \Omega_m = 0.287$, implemented as WMAP9 in the {\tt astropy.cosmology}
package (Astropy Collaboration 2013; 2018).


\begin{figure*}[t]
\includegraphics[width=8.8cm]{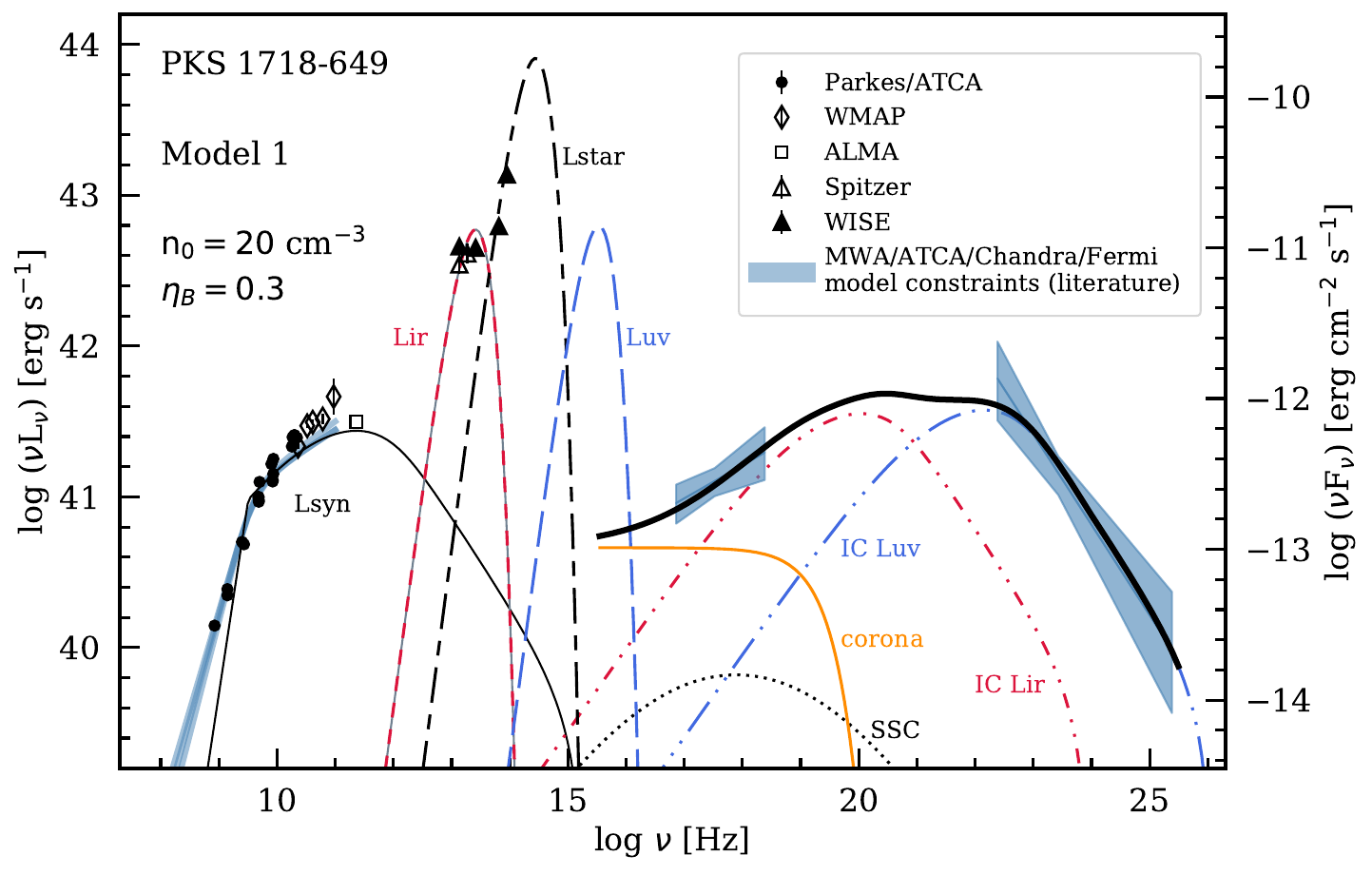}
\includegraphics[width=8.8cm]{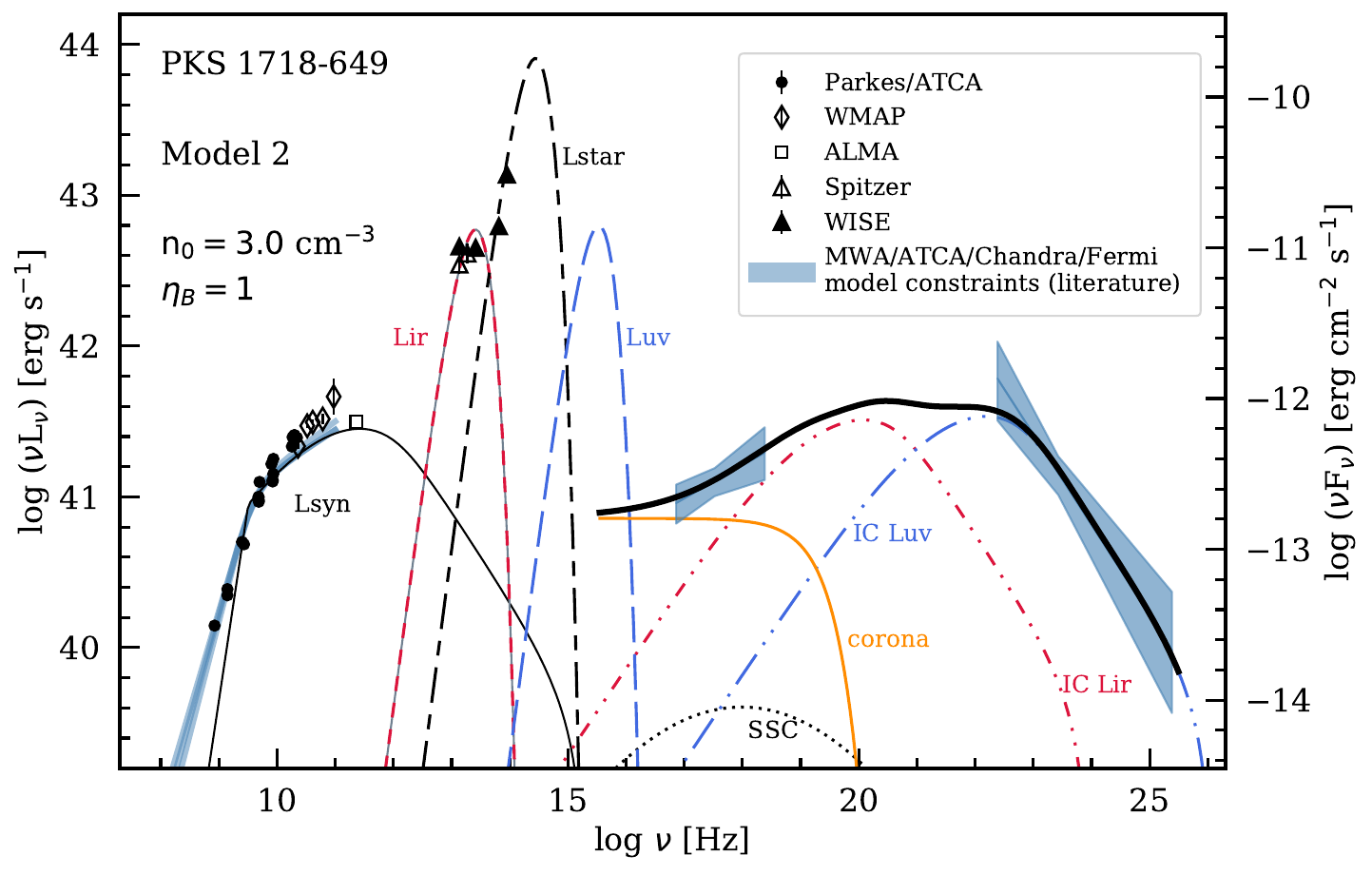}
\caption{Observed broadband spectral energy distribution of PKS\,1718$-$649
and theoretical models able to satisfactorily describe the data. The models
are characterized with a moderate departure of the magnetic field from equipartition,
$\eta_B = 0.3$ (Model 1, left) and $\eta_B = 1.0$ (Model 2, right). No formal
fitting was performed. Model parameters are listed in Tables~\ref{tab:sed_fixed}
and \ref{tab:sed}.
Circles -- radio data from Mauch et al. (2003), Maccagni et al. (2014),
Tingay et al. (2003), Bolton et al. (1975), Massardi et al. (2008),
Gregory et al. (1994), Murphy et al. (2010), Wright \& Otrupcek (1990),
Healey et al. (2007), Ojha et al. (2010), Sadler et al. (2006), Ricci et al. (2006).
Diamonds -- 9-year WMAP catalog data from Gold et al. (2011).
Square -- ALMA measurement from Maccagni et al. (2018).
Open triangles -- Spitzer measurements from Willett et al. (2010).
Filled triangles -- WISE measurements from Cutri et al. (2013; flux within 8\arcsec).
Semi-transparent blue lines in the $8 < \log \left( \nu / {\rm Hz} \right) < 11$ range
-- MWA/ATCA radio model constraints (Tingay et al. 2015). Butterfly regions -- 
\chandra\ and {\it XMM}-Newton model constraints on the intrinsic unabsorbed
power-law emission (Beuchert et al. 2018); \fermi/LAT $1\sigma$ model constraints
(Principe et al. 2021).
Broadband model components are as follows: self-absorbed synchrotron radiation
(thin black solid line), three black-body components representing the infra-red
(short-dashed), starlight (long-short-dashed) and accretion disk photon fields
(long-dashed), and their corresponding IC components originating from a single radio lobe. The short-dashed line
represents the $f_{\rm IR}$ corresponding to the black body
component normalized to $L_{\rm IR}$ at the $\nu_{\rm IR}$ frequency.
The solid orange line illustrates the contribution of
a low-luminosity X-ray nuclear emission (a weak X-ray corona or an ADAF-type emission).
The thick black line represent the sum of the IC components and the additional X-ray emission.}
\label{fig:sed}
\end{figure*}

\section{Multiwavelength data of \pks}
\label{sec:obs}

In this section, we summarize the multiwavelength observations of \pks.
We show the observed SED
of the source in Figure~\ref{fig:sed}.
We use the observational constraints in Section~\ref{sec:res}
to differentiate among various models of the broadband SED for \pks.

\subsection{Radio and submillimeter}

\pks\ has been thoroughly studied in the radio band with
the Parkes telescope (Bolton \& Butler 1975; Gregory et al. 1994),
Very Large Array (VLA; Healey et al. 2007),
Australia Telescope Compact Array (ATCA; Wright \& Otrupcek 1990;
Ver\'on-Cetty et al. 1995; Tingay et al. 1997, 2015;
Tingay \& de Kool 2003; Ricci et al. 2006; Sadler et al. 2006; 
Massardi et al. 2008; Murphy et al. 2010; Maccagni et al. 2014),
Murchison Widefield Array (MWA; Tingay et al. 2015). The source has been 
detected with the Wilkinson Microwave Anisotropy Probe (WMAP; Bennett et al. 2003; 
Chen \& Wright 2009; Giommi et al. 2009; Massardi et al. 2009;
Wright et al. 2009; Gold et al. 2011) and 
a continuum flux measurement of 0.304\,Jy at 230\,GHz
has been obtained with the Atacama Large Millimeter/submillimeter Array
(ALMA, Maccagni et al. 2018).
Representative radio to submillimeter flux measurements
are collected in Figure~\ref{fig:sed}.

The Very-Long-Baseline Interferometry (VLBI) observations at 4.8\,GHz
from ground and space revealed a compact double-sided structure
with separation of $\sim 7$\,mas, corresponding to the projected
linear size $LS \sim 2$\,pc at the redshift of the source, assuming
the orientation of the lobes in the place of the sky
(the Southern Hemisphere VLBI Experiment, SHEVE, Tingay et al. 1997;
the Highly Advances Laboratory for Communications and Astronomy,
HALCA, Tingay et al. 2002; 
the Long Baseline Array, LBA, Angioni et al. 2019)
and no apparent radio emission from the core (Tingay et al. 2002).
Multi-epoch radio monitoring allowed the
derivation of the hot spot advance velocity, $v_h/c \lesssim 0.07$
(Giroletti \& Polatidis 2006), which implied that the kinematic age
of the radio source in \pks\ is $\sim 100$ years;
see also Angioni et al. (2019) who found $v_h/c = 0.13 \pm 0.06$
and age of $70 \pm 30$ years.
Recent observations of \pks\ with ATCA and MWA allowed Tingay et al. 
(2015) to conclude that the radio data of the source are best modeled
with an inhomogeneous free-free absorption model (Bicknell et al. 1997).
The best fitting models derived by Tingay et al. (2015) for
their three observing runs are plotted in Figure~\ref{fig:sed}.

\subsection{Infrared and optical/ultraviolet} 

The source has been detected in the mid-infrared (MIR) band with 
the Wide-field Infrared Survey Explorer (WISE);
we include in Figure~\ref{fig:sed} the 4-band fluxes from the
AllWISE source catalog published by Cutri et al. (2013).
Willett et al.\ (2010) reported on {\it Spitzer} observations
of \pks.\  A $5.2 - 38$\,$\mu$m MIR spectrum of the source has been
obtained, and peakup
fluxes of $F(16{\rm \mu m}) = 49$\,mJy and $F({\rm 22\mu m}) = 57$\,mJy have been
measured with 15\% measurement error (see Figure~\ref{fig:sed}).

In addition, the authors detected a spectrally resolved
[OIV] 25.8\,$\mu$m line, which allowed them to estimate the mass of the black hole in \pks,
using the relation by Dasyra et al. (2008), $\log(M/M_{\odot}) = 8.62\pm0.45$,
consistent with the mass they derived using the L$_{\rm bulge}$-M$_{\rm BH}$ relation of
Bentz et al. (2009), $\log(M^{\rm bulge}_{\rm BH} / M_\odot) = 8.48$.
The MIR view of \pks\ revealed a moderately dusty environment and a
low star-formation rate in the host galaxy. The authors argue that a recent merger
triggered the AGN activity, but also stripped the star-forming gas from the galaxy.

Veron-Cetty et al. (1995) argued that the host of \pks\ resembles a high luminosity
elliptical galaxy with a faint outer spiral structure, which most likely originated
in a merger involving a gas-rich spiral in the process of forming an
elliptical. Optical spectroscopy of \pks's host galaxy, NGC 6328, was
presented in Filippenko (1985). The subtraction of an elliptical galaxy
template revealed a weak non-stellar power-law component classified as a LINER AGN,
which was found to contribute approximately half the strength of starlight near 3200\,\AA. 

\subsection{High energies}

\pks\ was observed for the first time in X-rays with \chandra\ in 2011 for 5\,ks
as part of our CSO X-ray survey (Siemiginowska et al. 2016). This initial
observation revealed that the X-ray spectrum of the point source can be described
by an absorbed power-law model with the photon index $\Gamma=1.6\pm0.2$ and
the intrinsic equivalent hydrogen column density $N_H(z) = (0.8\pm0.7) \times 10^{21}$\,cm$^{-2}$,
and that it is embedded in diffuse X-ray emission.
\pks\ was then re-observed with
\chandra\ in 2014 for a total time of 50\,ks and with {\it XMM}-Newton in 2017 for 20\,ks.
A simultaneous fit to these multi-epoch data allowed Beuchert et al. (2018) to 
detect a presence of non-variable emissions due to photoionized and
collisionally ionized plasmas; the former was explained as due to nuclear irradiation and
the latter as due to supernovae activity in the host galaxy.
They constrained the photon index to $\Gamma = 1.78^{+0.10}_{-0.09}$, and
found variability on the timescale of years in the normalization of the power law
emission by a factor of
up to $\sim$2.5, and in $N_H(z)$ in the
$(3-7) \times 10^{21}$\,cm$^{-2}$ range. Despite
its modest $N_H(z)$, the radio properties of \pks, such as its linear
size and power at 5\,GHz, place it on a low radio power extension
of a track occupied by CSOs with $N_H(z) > 10^{23}$\,cm$^{-2}$
in the radio size vs. radio luminosity vs. $N_H(z)$ diagram (Sobolewska et al. 2019a).
Based on this diagram, the X-ray obscured
CSOs appear to have smaller radio sizes, perhaps due to confinement by the environment,
compared to the X-ray unobscured CSOs with the same 5\,GHz radio power.
Alternatively, the X-ray obscured CSOs can be
seen as more radio loud compared to the X-ray unobscured CSOs with the same
linear radio size (see Sobolewska et al. 2019a for details).
Thus, the X-ray absorption and radio properties of \pks\ make it a particularly
interesting target for understanding the impact of the environment on the initial
radio source evolution.

PKS\,1718$-$649 was the first established CSO detected in the $\gamma$-ray band (Migliori et al. 2016).
The reported \fermi/LAT test statistic was $TS = 36$ ($\sigma \gtrsim 5$) 
band and $TS = 18.5$ ($\sigma = 4.3$) in the $0.1 - 100$\,GeV and $0.2 - 100$\,GeV band,
respectively.
The source has been included in the 4th $\fermi$/LAT catalog (Abdollahi et al. 2020),
and its $\gamma$-ray properties were revised by Principe et al. (2021), who found the 
$0.1 - 100$\,GeV $\gamma$-ray flux and slope of
$F_{\rm LAT} = (5.3\pm1.5)\times10^{-9}$\,ph\,cm$^{-2}$\,s$^{-1}$ and
$\Gamma_{\rm LAT} = 2.6\pm0.14$, respectively. 
No significant year-to-year or shorter $\gamma$-ray variability was detected from the source.

The confidence regions representing the
intrinsic X-ray and $\gamma$-ray power law emissions of the source are shown 
in Figure~\ref{fig:sed}. We adopted the $\gamma$-ray power-law parameters
as in Principe et al. (2021). In the X-ray band we use the photon
index and its error, and a mean X-ray power-law normalization and its mean
error estimated based on the X-ray fits presented in Beuchert et al. (2018).
This means that we model only the intrinsic power-law X-ray component arising
near the nucleus, and not the extended X-ray emission.

\section{Broadband model}
\label{sec:model}

We model the broadband SED of \pks\ with the dynamical model of Stawarz et al. (2008)
to investigate the origin of its high-energy (X/$\gamma$-ray) emission.
In this model, a set of equations originally considered by Begelman \& Cioffi (1989)
to describe classical doubles expanding in an ambient medium is employed to
characterize the evolution of compact sources. In the framework of this model,
a relativistic jet with kinetic
power $L_{\rm j}$ propagates in the innermost parts of the host galaxy with a constant
velocity, $v_h$, into a uniform gaseous
medium of constant density $\rho = m_{\rm p} n_{\rm 0}$.
The momentum flux of the relativistic jet is balanced by the ram pressure of the ambient
medium spread over some constant area $A_{\rm h}$, $L_{\rm j}/c = m_{\rm p} n_{\rm  0} v_{\rm h}^2 A_{\rm h}$.
Ultra-relativistic electrons with an initial energy distribution $Q(\gamma)$ and Lorenz factors
$\gamma_{\rm min}<\gamma<\gamma_{\rm max}$ are injected from the terminal hot spots 
of the jet into the expanding lobes. The electron population of the lobes
undergoes
adiabatic and radiative cooling in the course of the source growth from
an initial size $r_0 =  (A_h \pi^{-1} l_c^{-2}) LS$ to $LS$.
Transverse expansion of the source is governed by a scaling law $l_c(t) \propto t^{1/2}$
reproducing the initial, ballistic phase of the jet propagation
(e.g. Kawakatu \& Kino 2006). The lobe's electrons inverse-Compton scatter all the
ambient low-energy photon populations, which include synchrotron photons, infrared (torus)
photons, galaxy starlight, and the ultraviolet radiation of an accretion flow.
We approximate the infrared, optical and UV spectra as blackbody components for
the purpose of evaluating the IC radiation of the lobes.
The synchrotron radiation is described with a synchrotron self-absorbed model.
We refer the readers
to Stawarz et al. (2008) and Ostorero et al. (2010) for further details of the model.

The model has many parameters. However, high quality observations of \pks\
across the electromagnetic spectrum allowed us to put constraints on the 
majority of them (Table~\ref{tab:sed_fixed}). We use the linear size of
the radio source, $LS = 2$\,pc, and the hotspot separation velocity, $v_{\rm h}
= 0.07c$ reported by Giroletti \& Polatidis (2009; see also Angioni et al. 2019 for a
more recent measurement).
We note that our model assumes a source with perfectly symmetric morphology.
Therefore, the model describes the evolution of one of the two lobes only,
with the core-hotspot distance taken as $LS^{\prime} = LS/2$, and the separation
velocity of the hotspot from the core taken as $v^{\prime}_h = v_{\rm h}/2$.
The luminosity of the modeled lobe
is then multiplied by a factor of two to  be compared with the observed lobes'
luminosity.

We fix the radio 
turnover frequency at the average of the values derived by Tingay et 
al. (2015) from ATCA observations of the source.
We normalize the infrared blackbody component using the {\it Spitzer} flux
at 16\,$\mu$m (corresponding to frequency $\nu_{\rm IR}$).
Given the dominance of starlight at optical frequencies in \pks\ and related
uncertainties in the galaxy-AGN decomposition performed
by Filippenko et al. (1985), for simplicity we fix the V-band and UV luminosities 
at $\nu_{\rm star}$ and $\nu_{UV}$ ($L_{\rm star}$, $L_{\rm UV}$; see
Table~\ref{tab:sed_fixed}) at values that result in both blackbody components
having comparable fluxes at 3200\,\AA, corresponding to
$m_{\rm AB} \sim 18$, where $m_{\rm AB} = -2.5 \log ( f_{\nu} ) - 48.60$,
and $f_{\nu}$ is in erg\,s$^{-1}$\,cm$^{-2}$\,Hz.
We note that the blackbody component describing the visible light in our
model matches rather well the W1 and W2 WISE
measurements of Cutri et al. (2013).

The model assumes that the lobe electrons provide the bulk of the lobe pressure.
The electron energy density is $U_e = \eta_E p$ with $\eta_E \lesssim 3$.
The magnetic field is given by $B = (8 \pi \eta_B p)^{1/2}$, with
$\eta_B = U_B / p < 3$, where $U_B$ denotes the magnetic field energy density.
In this paper, we follow Stawarz et al. (2008) and we choose $\eta_E = 3$,
and we assume that the lobe electrons are in rough equipartition with
the magnetic field and protons (e.g.  Orienti \& Dallacasa 2008, and references therein).
We consider $\eta_B = 0.3$, which implies that the ratio of the model magnetic field
energy density to the equipartition magnetic field energy density
$U_B / U^{\rm eq}_B = 0.1$,
or $B/B^{\rm eq} \sim 0.3$;
$\eta_B = 1$ ($U_B / U^{\rm eq}_B = 1/3$, $B/B^{\rm eq} \sim 0.6$);
and $\eta_B = 3$ ($U_B = U^{\rm eq}_B$, $B = B^{\rm eq}$).

As discussed in Stawarz et al. (2008), the likely shapes of the initial electron
distribution injected into the radio lobes include
a power-law function, $Q(\gamma) \propto \gamma^{-s}$, or a broken power-law 
function with the slope of the distribution changing from $s_1$ to $s_2$ at a 
given Lorentz factor $\gamma_{\rm b}$. We test
both possibilities.
We choose $\gamma_{\rm min} = 1$ and $\gamma_{\max} = 100\,{\rm m_p/m_e}$.

We note that while Tingay et al. (2015) demonstrated that a proper description of
the low-energy radio SED should include inhomogenous free-free absorption processes,
we use the standard synchrotron self-absorbed spectrum. This choice does not
affect the final high-energy shape of the $\gamma$-ray model SED because
this regime is dominated by the IC scattering off high-energy electrons.

Given the above assumptions and observational constraints, we are left with only
a handful of free model parameters:
(i) the density of the ambient medium in which the lobes expand, $n_0$;
(ii) the jet kinetic power, $L_{\rm j}$; (iii) the
parameters of the electron energy distribution $Q(\gamma)$:
the slope $s$ if $Q(\gamma)$ is described by a single power function,
or the slopes $s_1$ and $s_2$ of the lower- and higher-energy parts
and the Lorentz factor corresponding to the break, $\gamma_{\rm b}$,
if $Q(\gamma)$ is described by a broken power-law function.

\begin{table}[t]
{\scriptsize
\noindent
\caption[]{\label{tab:sed_fixed} Fixed parameters of the broadband SED model}
\begin{center}
\begin{tabular}{lllll}
\hline\hline
 Description & Symbol & Value & Unit & Refs.$^a$ \\ \hline
 \multicolumn{5}{c}{Geometry of the radio source} \\ \hline
Linear size                      & LS                 & 2    & pc  & (1)$^b$ \\
Hotspot sep. velocity            & $v_{\rm h}$        & 0.07 & $c$ & (1)$^b$ \\ \hline
\multicolumn{5}{c}{Characteristic frequencies and luminosities} \\ \hline
Radio turnover freq.             & $\nu_{\rm peak}$   & 3.25 & GHz & (2)$^c$ \\
IR band ref. freq.               & $\nu_{\rm IR}$     & 1.87 & 10$^{13}$\,Hz & $\dots$ \\
$\nu L_{\nu}$ @ $\nu_{\rm IR}$   & $L_{\rm IR}$       & 8.05 & 10$^{42}$\,erg\,s$^{-1}$ & (3) \\
Visible band ref. freq.          & $\nu_{\rm star}$   & 2.0  & 10$^{14}$\,Hz & $\dots$ \\
$\nu L_{\nu}$ @ $\nu_{\rm star}$ & $L_{\rm star}$     & 1.1  & 10$^{44}$\,erg\,s$^{-1}$ & (4) \\
UV band ref. freq.               & $\nu_{\rm UV}$     & 2.45 & 10$^{15}$\,Hz & $\dots$ \\
$\nu L_{\nu}$ @ $\nu_{\rm UV}$   & $L_{\rm UV}$       & 8.5  & 10$^{42}$\,erg\,s$^{-1}$ & (5) \\ \hline
\hline
\end{tabular}
\end{center}
\smallskip Notes:\\
$^a$ References: (1) Giroletti \& Polatidis (2009); (2) Tingay et al. (2015);
(3) Willett et al. (2010); (4) Filippenko (1985).
$^b$ We use $LS^{\prime} \simeq LS/2 = 1$\,pc and $v_{\rm h}^{\prime} \simeq v_{\rm h}/2 = 0.035c$
for core-hotspot linear size and separation velocity (c.f. Ostorero et al. 2010).
$^c$ The average of the $\nu_{\rm peak}$ values reported in Tingay et al. (2015).
}
\end{table}

\begin{table*}[t]
{\scriptsize
\noindent
\caption[]{\label{tab:sed} Variable parameters of the broadband SED model}
\begin{center}
\begin{tabular}{llllll}
\hline\hline
 Description & Symbol & Model 1 & Model 2 & Unit & Refs.$^a$ \\
\hline
\multicolumn{6}{c}{Injected electron population $Q(\gamma)$ and jet kinetic power} \\ \hline
Low-energy slope                 & $\rm s_1$          & 1.9  & 1.85 &     & ($\star$)$^b$ \\
High-energy slope                & $\rm s_2$          & 3.2  & 3.2  &     & ($\star$) \\
Lorentz factor (min)             & $\gamma_{\rm min}$ & 1    & 1    &     & (1) \\
Lorentz factor (break)           & $\gamma_{\rm b}$   & 3    & 3    & $\rm m_p/m_e$ & ($\star$) \\
Lorentz factor (max)             & $\gamma_{\rm max}$ & 100  & 100  & $\rm m_p/m_e$ & (2) \\
Jet kinetic power                & $L_{\rm j}$        & 2.2  & 1.7  & 10$^{42}$\,erg\,s$^{-1}$ & ($\star$) \\ \hline
\multicolumn{6}{c}{Environment} \\ \hline
ISM density                      & $n_0$              & 20  & 3 & cm$^{-3}$ & ($\star$) \\
Electrons                        & $\eta_E$           & 3   & 3 &  & (1)$^c$ \\
Magnetic field                   & $\eta_B$           & 0.3 & 1 &  & (1, $\star$)$^c$ \\
\hline
\multicolumn{6}{c}{Additional power law X-ray emission} \\ \hline
Photon index                     & $\Gamma$ & 2.0 & 2.0 &  & ($\star$) \\
Luminosity (2-10 keV) & $L_{\rm 2-10\,keV}$ & 6.6 & 8.6 & 10$^{40}$\,erg\,s$^{-1}$  & ($\star$) \\
\hline\hline
\end{tabular}
\end{center}
\smallskip Notes:\\
$^a$ References: ($\star$) This work; (1) Stawarz et al. (2008); (2) Ostorero et al. (2010).
$^b$ The radio spectral slope $\alpha\simeq 0.7$ (Tingay et al. 2015) suggests
${\rm s_1} = 2\alpha +1 \simeq 2.4$ for the {\it evolved} electron distribution.
$^c$ Electron and magnetic field energy densities are parametrized as $U_E = \eta_E p$ and
$U_B = \eta_B p$, where $p$ stands for the expanding cocoon's internal pressure.
}
\end{table*}


\section{Results}
\label{sec:res}

The expanding radio lobe model can successfully reproduce the bulk of the high-energy emission
of \pks. The IC scattering of the UV photons
(presumably from an inner accretion flow) can account for the \fermi/LAT observational constraints,
while the IC scattering of the infrared photons detected from
the direction of \pks, presumably due to the emission of a dusty environment in
the galactic center, contributes to the source's X-ray emission.
We found that the contributions to the high energy emission of \pks\
coming from the IC scattering of the optical (galaxy) and synchrotron photons are negligible.
The variable parameters of our final models are collected in 
Table~\ref{tab:sed} and the model SEDs are plotted in Figure~\ref{fig:sed}.
Below we describe in detail our modeling rationale and results.

Models with $Q(\gamma)$ in the form of a single power-law function
were found to substantially overestimate the $\gamma$-ray emission and they failed
to reproduce the observed $\gamma$-ray photon index. Thus, we concluded that 
the $\fermi$/LAT constraints require that the energy distribution of
the electrons injected into the lobes of this source has a broken
power-law form.

Beuchert et al. (2018) measured the intrinsic equivalent hydrogen absorbing
column density from the direction of \pks\ and found that it varies in the
$N_H \sim (3-7) \times 10^{21}$\,cm$^{-2}$ range. The radii
of the regions used in that work to extract the
\chandra\ (14\,$\arcsec$) and {\it XMM}-Newton (40\,$\arcsec$) energy spectra 
correspond to $\sim 4$ and $\sim 12$\,kpc, respectively, at the redshift of the source.
Thus, the location of
the intrinsic matter obscuring the nuclear X-rays cannot be determined on a parsec
scale in \pks\ based on the modeling of Beuchert et al. (2018).
We checked that with an extraction region size of 1.5\arcsec\
($\sim 440$\,pc at the redshift of the source),
approaching the spatial resolution of \chandra, the \chandra\ data of the source
(ObsIDs 16070 and 16623) are still consistent with intrinsic
$N_H \sim (2-3) \times 10^{21}$\,cm$^{-2}$.
An intrinsic column density of the order of $3 \times 10^{21}$\,cm$^{-2}$
implies a particle density within 440\,kpc of $n_0 \sim 2.4$\,cm$^{-3}$.

However, we found that models with $n_0$ of this order
underestimate the ALMA and WMAP measurements for $\eta_B = 0.3$.
In particular, the continuum flux at 290\,GHz measured by
Maccagni et al. (2018) is underestimated by $\gtrsim 30$\%
(see Model A1 in Appendix;
Figure~\ref{fig:sed_appendix} and Table~\ref{tab:sed_appendix}).
A density an order of magnitude higher, $n_0 = 20$\,cm$^{-2}$,
is required to fully account for the observed radio-to-submillimeter
band in the broadband SED model of \pks\ (Model 1, Figure~\ref{fig:sed}).
This value of $n_0$ implies that the X-ray $N_H$ measured by Beuchert et al. (2018)
is distributed uniformly within the central 50-100\,pc.  

Alternatively, we found that for $n_0 = 3$\,cm$^{-3}$ a good match of
the model with the data can be obtained by setting $\eta_B = 1$
(Model 2, Figure~\ref{fig:sed}), 
which brings the physical conditions closer to equipartition than the case with
$\eta_B = 0.3$. In the equipartition case, $\eta_B = 3$, the soft
$\gamma$-ray band becomes underestimated by the model,
unless the density is set to $n_0 \lesssim 1$\,cm$^{-3}$
(Model A2 in the Appendix).

The $\gamma$-ray spectrum (both the photon index and normalization) of \pks\
can be accounted for satisfactorily by Models 1 and 2
with $s_1 \sim 1.9$, $s_2 = 3.2$,
$\gamma_{\rm b} = 3$\,${\rm m_p/m_e}$,
and jet kinetic power $L_j = (1.7 - 2.2) \times 10^{42}$\,erg\,s$^{-1}$
(Table~\ref{tab:sed}), and the remaining model 
parameters fixed at the values derived from observations or assumed as described in
Section~\ref{sec:model}.
However, these models returned an X-ray photon index that was
significantly harder than the observed one, and underestimated the observed X-ray
emission. Given that \pks\ contains a LINER type AGN, it is possible that
a low luminosity nuclear emission contributes to the X-ray emission of the source.
We modeled this with a cutoff power law function with
$\Gamma = 2.0$, $E_{\rm cutoff} = 100$\,keV,
and 2-10\,keV luminosity of $(6.6 - 8.6) \times 10^{40}$\,erg\,s$^{-1}$. We
found that a sum of such a power-law component and IC scattering of the infrared
photons in the radio lobes is able to fully explain the observed X-ray emission of \pks.

\section{Discussion}
\label{sec:disc}

We explored the applicability of the expanding radio lobe model (Stawarz et al. 2008)
to the broadband radio-to-$\gamma$-ray SED of one of the youngest,
most compact, and nearest symmetric radio sources known to date, \pks.
Our modeling allowed us to uncover possible mechanisms responsible for the high-energy
emission of the source and constrain interesting physical parameters of the
source, such  as the spectrum of the electrons injected from the hot spots into
the lobes, and the jet kinetic power.

In general, our results suggest a moderate departure,
within one order of magnitude,
of the magnetic field strength in the radio lobes of \pks\
from equipartition. We note that such a departure from equipartition is supported
by observations of other radio sources (e.g. Ineson et al. 2017, Croston et al. 2018;
see, however, Orienti \& Dallacasa 2008).

An equally good match of Models 1 and 2 with the data indicates that 
there exists a degeneracy in the expanding radio lobe model between the density of
the ambient medium and the degree to which the magnetic field in the
lobes deviates from equipartition, which cannot be broken with the current
data. However, the remaining variable model parameters are either virtually
identical in Models 1 and 2 (parameters of the injected electron population and
the photon index of the additional X-ray component) or within 30\% from
each other (the jet power and the $2 - 10$\,keV luminosity of the additional
X-ray component), as detailed in Table~\ref{tab:sed} and discussed in Sections
5.1--5.5 below.

In addition, we noted that the strict equipartition case requires that
$n_0 \lesssim 1$ (Model A2 in the Appendix), which is in conflict with the X-ray 
measurements (Beuchert et al. 2018; this work),
unless the density distribution in the host of \pks\ follows a profile such that
$n_0 \lesssim 1$ on the few parsec scale comparable with the separation between
the radio lobes, and increases to $n_0 \gtrsim 3$\,cm$^{-3}$ on the few hundred 
kiloparsec scale corresponding to the 1.5$\arcsec$ extraction region resolved 
with  \chandra.

We show in the Appendix that models with high $\eta_B$ and high
density (Models A3 and A4) have difficulties in accounting for the high-energy emission in
\pks: the X-ray band becomes dominated by an additional
X-ray component in these models, and the soft $\gamma$-ray emission is underestimated
(given the {\it Fermi}/LAT 1$\sigma$ confidence level model constraints
by Principe et al. 2021).

\subsection{Origin of the $\gamma$-ray emission}
We found that the properties of the $\gamma$-ray emission observed from \pks\ put
strong constraints on
the electron distribution $Q(\gamma)$ injected into the radio lobes. It is required
that this distribution has a broken power-law shape, and it is
characterized by a break energy $\gamma_{\rm b} = 3\,{\rm m_p/m_e}$ and a high-energy
slope $s_2 = 3.2$. The radio and submillimeter data imply that the lower-energy segment
($\gamma < \gamma_{\rm b}$) of $Q(\gamma)$ has
index $s_1 \sim 1.9$. The lobes expand in a medium with density
$n_0$ in the $3 - 20$\,cm$^{-3}$ range for $\eta_B$ in the $1 - 0.3$ range, i.e.
the higher the density the lower the $\eta_B$ parameter.

The electron population
of the radio lobes evolves during the lobes' expansion
due to the adiabatic and radiative cooling effects (Stawarz et al. 2008).
In the case of a
broken power-law injection, the electron spectral continuum steepens at
$\gamma > \gamma_{\rm cr} = 200\,\eta_B^{-1} L_{j, 45}^{-1/2}$ when compared
to the injected one, where $L_{\rm j, 45} \equiv L_{\rm j} / 10^{45}$\,erg\,s$^{-1}$.
In our model solutions for \pks, $\gamma_{\rm cr}$ corresponds to
$\sim 7.7\,{\rm m_p / m_e}$ ($\sim 8.8\,{\rm m_p / m_e}$)
or $\sim 2.6\,\gamma_{\rm b}$ ($\sim 2.9\,\gamma_{\rm b}$) for $\eta_B = 0.3$
($\eta_B = 1.0$).
The $s_1$ index is close to the canonical spectrum generated by
diffusive (first-order Fermi) shock acceleration and comparable with the
low-energy slopes derived by Ostorero et al. (2010) for
a sample of eleven GPS/CSO galaxies known as X-ray emitters up to 2008.
Interestingly, ambient density $n_0 = 20$\,cm$^{-3}$ in Model 1 with $\eta_B=0.3$
is of the same order as the mean density found 
by Mukherjee et al. (2016; 2017) who fitted the probability density function of
the simulated density of an ISM evolving in the presence
of an expanding jet
with jet head $\lesssim 1$\,kpc using a modified lognormal function proposed
by Hopkins (2013; see also Zovaro et al. 2019).

\subsection{Origin of the X-ray emission}
We showed that, while the $\gamma$-ray emission in \pks\ is consistent with IC scattering
of the UV photons from an accretion flow off energetic electrons in
the radio lobes, the origin of the X-ray emission is more elusive.
We showed that the IC scattering of the infrared ambient photons, likely
due to the emission of the dusty environment of \pks\ resolved with VLT and ALMA
(Maccagni et al. 2016; 2018), contributes to the X-ray emission of the source.
However, its photon index is harder than that derived from the observations.
As a result, the model was found to underestimate the soft X-ray emission of the source.
We proposed that an additional component is required in order to
fully explain the observed intrinsic X-ray emission, and we
modeled it with a power-law function with a photon index
$\Gamma = 2.0$, a 2-10\,keV luminosity
$L_{\rm 2-10\,keV} = (6.6 - 8.6) \times 10^{40}$\,erg\,s$^{-1}$ and
a high-energy cutoff at 100\,keV (even though the current data cannot confirm
or reject the presence of such a cutoff).
Interestingly, these parameters are typical
for LINER type AGNs studied in X-rays (e.g. Gonzalez-Martin et al. 2009).
This additional component may be associated with a weak X-ray corona or
a radiatively inefficient nuclear emission (e.g. an ADAF; Ichimaru 1977;
Narayan \& Yi 1994; 1995a; 1995b; Abramowicz et al. 1995; Chen et al. 1995).
The sum of the IC X-ray emission from the radio lobes and low luminosity nucleus
agrees within the errorbars with the X-ray constraints derived by Beuchert et al.
(2018) using a single power-law function and interpreted as due to an X-ray corona
(see Figure~\ref{fig:sed}).

However, we stress that the relative contributions of a low luminosity nucleus
and IC scattering of the infrared photons in the radio lobes are difficult to constrain
with the current data. It is possible that only a fraction of the IR emission measured with
{\it Spitzer} intersects the expanding radio lobes, given the small size of the radio source in
\pks\ ($LS = 2$\,pc) and the complex structures in the innermost 15\,kpc of the host galaxy
of \pks. Indeed, resolved measurements with ALMA revealed a CO gas distributed
in a complex warped disk, forming a circumnuclear disk at $r \lesssim 700$\,pc,
and molecular clouds falling onto the central supermassive black hole at
$r \lesssim 75$\,pc (Maccagni et al. 2018). We estimated that if only half of the
infrared emission becomes IC scattered in the radio
lobes in Model 1 ($\eta_B = 0.3)$, then our modeling requires a nuclear power-law
X-ray component with $L_{\rm 2-10\,keV} = 5.5 \times 10^{40}$\,erg\,s$^{-1}$,
which is still well within the range of the 2-10\,keV luminosities reported in the
literature for the LINER-type galaxies. 

We note that our broadband model predicts a spectral hardening above 10\,keV
where the IC component
from the radio lobes dominates over the nuclear emission. This could be tested with 
observations in the hard X-ray band. In addition, the high energy SED of \pks\ peaks
in the MeV range, making \pks\ an ideal target for future MeV-band missions such as e.g.
AMEGO-X (Fleischhack \& Amego X Team 2022).

\subsection{Jet kinetic power and feedback}
The jet kinetic power resulting from Models 1 and 2,
$L_j = (1.7 - 2.2) \times 10^{42}$\,erg\,s$^{-1}$,
is in excellent agreement with the upper limit reported by Maccagni et al. (2014;
$L_j < 2 \times 10^{43}$\,erg\,s$^{-1}$).
This jet power is lower than the values found for 
powerful flat-spectrum radio quasars (FSRQs) and BL Lacertae sources
(e.g., Sambruna et al. 2006; Ghisellini et al. 2009), as well as for FRI/FRII radio 
galaxies
(e.g., Laing et al. 2002; Croston et al. 2008, 2009; Croston et al. 2018 and references 
therein; see also Xu et al. 2009).

The implied magnetic field intensity within the lobes is $\sim 4.3$\,mG
for $\eta_B = 0.3$ and $\sim 4.0$\,mG for $\eta_B = 1$, in agreement with 
expectations for compact young lobes.

We stress that the jet powers resulting from our preferred models
(Figure~\ref{fig:sed}, Table~\ref{tab:sed}) and models ${\rm A2 - A4}$
in the Appendix (Table~\ref{tab:sed_appendix})
vary at most by a factor of $\sim 3$,
indicating that our estimate of $L_j$ is robust.

Mukherjee et al. (2016; 2017) showed with numerical simulations
that the feedback of low-power jets is significant
because they are confined by the ISM for a longer time than their more powerful
counterparts. This affects the ISM density distribution and inhibits the star
formation. Indeed, Willett et al. (2010) reported a rather weak star formation
rate in \pks, $0.8 - 1.9$\,$M_{\odot}$\,yr$^{-1}$,
estimated by means of PAH signatures measured with {\it Spitzer}.
Interestingly, Mukherjee et al. (2017) argued that jets
with power $\lesssim 10^{43}$\,erg\,s$^{-1}$, such as those of \pks,
may be too weak to escape the ISM confinement, and too weakly pressurized to
prevent an infall of gas back into
the initially created central cavity.

\subsection{Mass accretion rate and gas mass reservoir}
We used the relation of Allen et al. (2006) to translate $L_j$
resulting from Models 1 and 2 into the Bondi accretion power,
$L_{\rm bondi} = (1.9 - 2.2) \times 10^{43}$\,erg\,s$^{-1}$.
Assuming an accretion efficiency $\epsilon = 0.1$, the Bondi accretion rate is
$\dot{M} = (0.003 - 0.004)\,{\rm M_\odot / yr}$.

Model 2 with $\eta_B = 1$ and $n_0 = 3$\,cm$^{-3}$, of the
order of that we found by modeling the innermost 1.5\arcsec\
X-ray region around \pks, allows for an estimation of a gas mass
reservoir within $\simeq 440$\,pc at $1.5 \times 10^7\,M_\odot$.

However, if the intrinsic $N_H$ measured in the X-ray band
by Beuchert et al. (2018) is indeed distributed uniformly over a surface area with a radius
$r = 50 - 100$\,pc, as suggested by Model 1 with $\eta_B = 0.3$ which favors an ambient
density of $n_0 = 20$\,cm$^{-3}$, then the gas mass reservoir
within this radius is of the order of $(0.3 - 1.3) \times 10^6\,M_\odot$,
providing an ample supply to feed the central black hole. 
The mass
accretion rate and the gas mass reservoir at such levels are compatible
with those reported by Maccagni et al. (2018), who inferred the presence
of cold clouds falling onto the central black hole within $\lesssim 75$\,pc
by studying tracers such as H$_{\rm I}$, H$_{\rm 2}$ and $^{\rm 12}$CO (2–1) with ALMA
and SINFONI (Maccagni et al. 2016; 2018).
They found an accretion rate $1.3 \times 10^{-3}$\,M$_{\odot}$\,yr$^{-1} \lesssim \dot{M}_{H_2} \lesssim 2.2$ \,M$_{\odot}$\,yr$^{-1}$, and a mass of the absorbing molecular clouds in the
$3 \times 10^2$\,M$_{\odot}$ - $5 \times 10^5$\,M$_{\odot}$ range.

The mass accretion rate that we found, expressed in terms of the Eddington
accretion rate, is
$\dot{M} = (4 - 5) \times 10^{-4}\,\dot{M}_{\rm Edd}$, assuming the mean of the
two black hole mass estimates in Willett et al. (2010).
On the other hand, the UV luminosity in our model, $L_{\rm UV} \simeq 8.5 \times 10^{42}$\,erg\,s$^{-1}$,
implies an accretion rate of the order of
$\dot{M} \simeq 2 \times 10^{-4}$\,$\dot{M}_{\rm Edd}$.
Both estimates are consistent with the LINER classification
of the active nucleus in \pks.

\subsection{Transverse expansion}
The transverse size and recent-day transverse expansion velocity
resulting from Models 1 and 2 are
$l_{\rm c} = (2.4 - 2.6)$\,pc and $v_{c} \sim 0.13\,c$,
respectively. They are larger than
their counterpart parameters along the core-hotspot direction (i.e., the core-hotspot
distance, ${\rm LS^{\prime}}$, and the core-hotspot separation velocity,
${v_{\rm h}^{\prime}}$).
Thus, the models suggest that the lobes of \pks, and other extremely compact radio sources,
may be more elongated in the transverse direction than in the direction of the hotspots.
Some evidence for this can be seen at least for the northern lobe on the 8.4\,GHz VLBI
radio maps recently presented by Angioni et al. (2019).
Since $l_c \propto {\rm LS}^{1/2}$ and $v_c \propto {\rm LS}^{-1/2}$,
the models predict that
eventually the transverse expansion will slow down: in \pks,
$v_c$ will be comparable to ${v_{\rm h}^{\prime}}$ once ${\rm LS^{\prime}}$ reaches
$\sim 10 - 15$\,pc, and the source will evolve to a state in which $l_c < LS^{\prime}$.

\section{Conclusions}
\label{sec:concl}

We demonstrated that the expanding radio lobe model by Stawarz et al. (2008) can explain the
high-energy emission in \pks, the first and one of only a few symmetric
young radio sources detected to date in the $\gamma$-ray band with {\it Fermi}/LAT, as being due to IC scattering of the IR and UV emission
off energetic electrons injected into the lobes from the hotspots,
assuming a rough equipartition between
the magnetic field and particles, and an additional contribution from a weak
X-ray corona or an ADAF at the luminosity level expected in LINER type AGNs.
Our results suggest that \pks\ is destined to evolve into a low-power
FRI type radio galaxy. Low power jets, like those in \pks, are
important for the jet/galaxy feedback process because they struggle
to propagate through the ISM on their way out from the host galaxy,
and interact with the gas in the host galaxies for a longer time
than their more powerful counterparts.

Based on our modeling of \pks, we
were able to estimate the magnetic field intensity within the lobes,
the shape of the distribution of the evolved electrons in the lobes,
the properties of the transverse lobe expansion,
the mass accretion rate, and the gas mass reservoir
available to feed the black hole.

The expanding radio lobe model has recently been considered
by Lister et al. (2020) in a discussion of
the high-energy SED of another $\gamma$-ray detected CSO, TXS\,0128+554. 
The authors reported that the observed {\it Fermi}/LAT flux of TXS\,0128+554
is three orders of magnitude higher than predicted $\gamma$-ray
emission from the lobes, and concluded that it most
likely originates in the inner jet/core region rather than in the lobes.
Furthermore, Sobolewska et al. (2019b) concluded that the expanding radio lobe model
for a set of model parameters considered by Ostorero et al. (2010) for OQ+208,
a CSO embedded in a cloud of matter with
an intrinsic absorbing column density of the order of 10$^{24}$\,cm$^{-2}$
(however, with no $\gamma$-ray detection to date),
appears to overestimate the level of the X-ray emission measured from
a joint modeling of {\it Chandra}, {\it XMM}-Newton and {\it NuSTAR} data.
Thus, it remains to be determined if the model of expanding radio lobes provides
a universal explanation of the X-ray and $\gamma$-ray emission of the Compact Symmetric
Objects and other GPS galaxies, or whether these sources form a heterogeneous population with
respect to the origin of their high-energy emission.

\begin{acknowledgments}
M.S. and A.S. were supported by NASA contract NAS8-03060 (Chandra X-ray Center).
M.S. acknowledges partial support by the NASA contracts 80NSSC18K1609
and 80NSSC19K1311. Partial support for this work was provided by the NASA
grants GO1-12145X, GO4-15099X.
L.O. acknowledges partial support from the INFN grant InDark and the
Italian Ministry of Education, University and Research (MIUR) under
the Departments of Excellence grant L.232/2016.
{\L}.S. was supported by the Polish National Science Center grant
2016/22/E/ST9/00061.
This research has made use
of NASA’s Astrophysics Data System Bibliographic Services. The authors thank
Rafa\l{} Moderski for making his numerical code of expanding radio lobes
available for this study, Mitchell Begelman for multiple discussions
on young radio sources, and the anonymous reviewer for insightful comments
on the manuscript.
\end{acknowledgments}

\appendix

\section{Supplementary model solutions}

\begin{figure*}[h]
\begin{center}
\includegraphics[width=8.8cm]{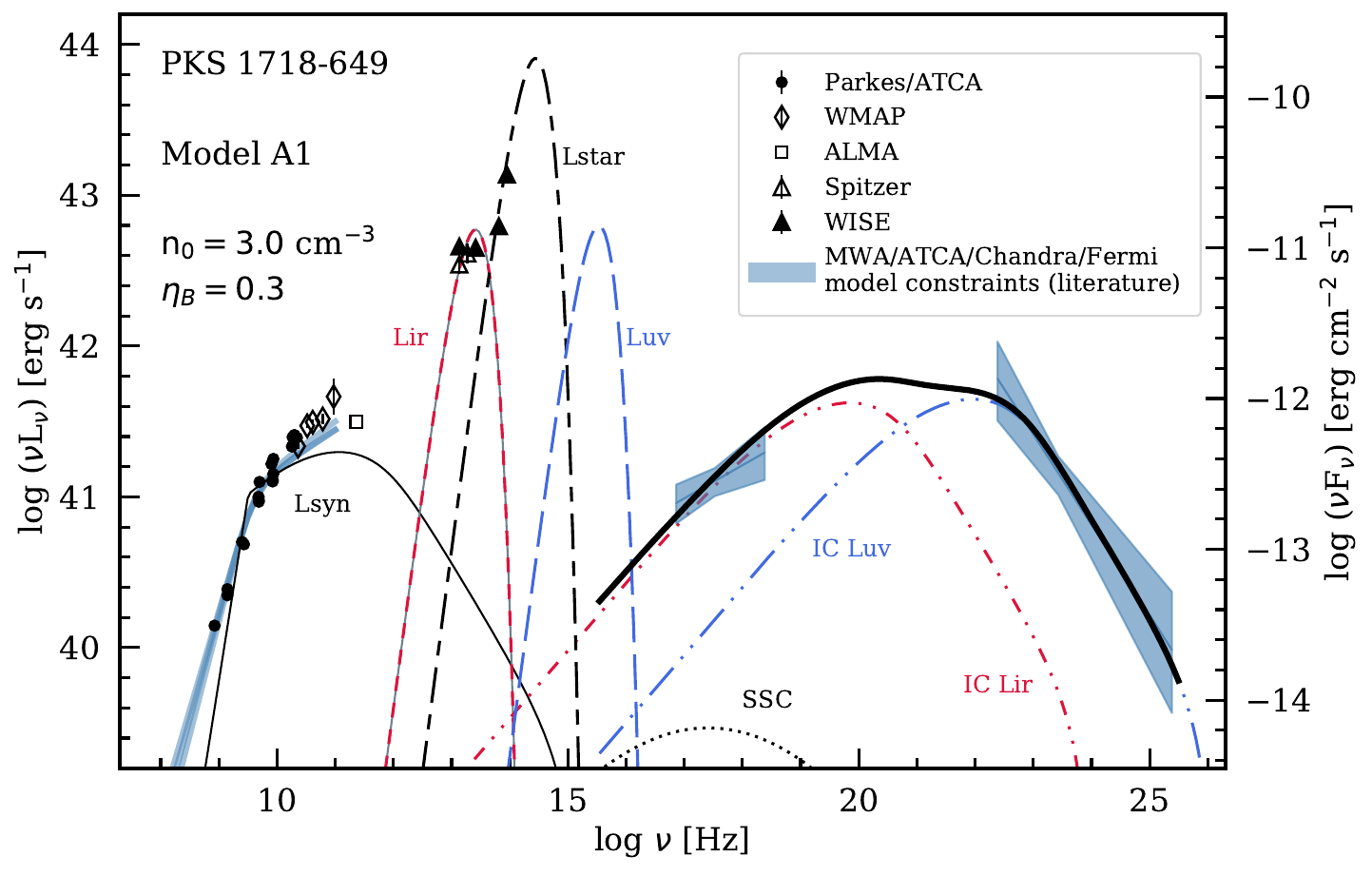}
\includegraphics[width=8.8cm]{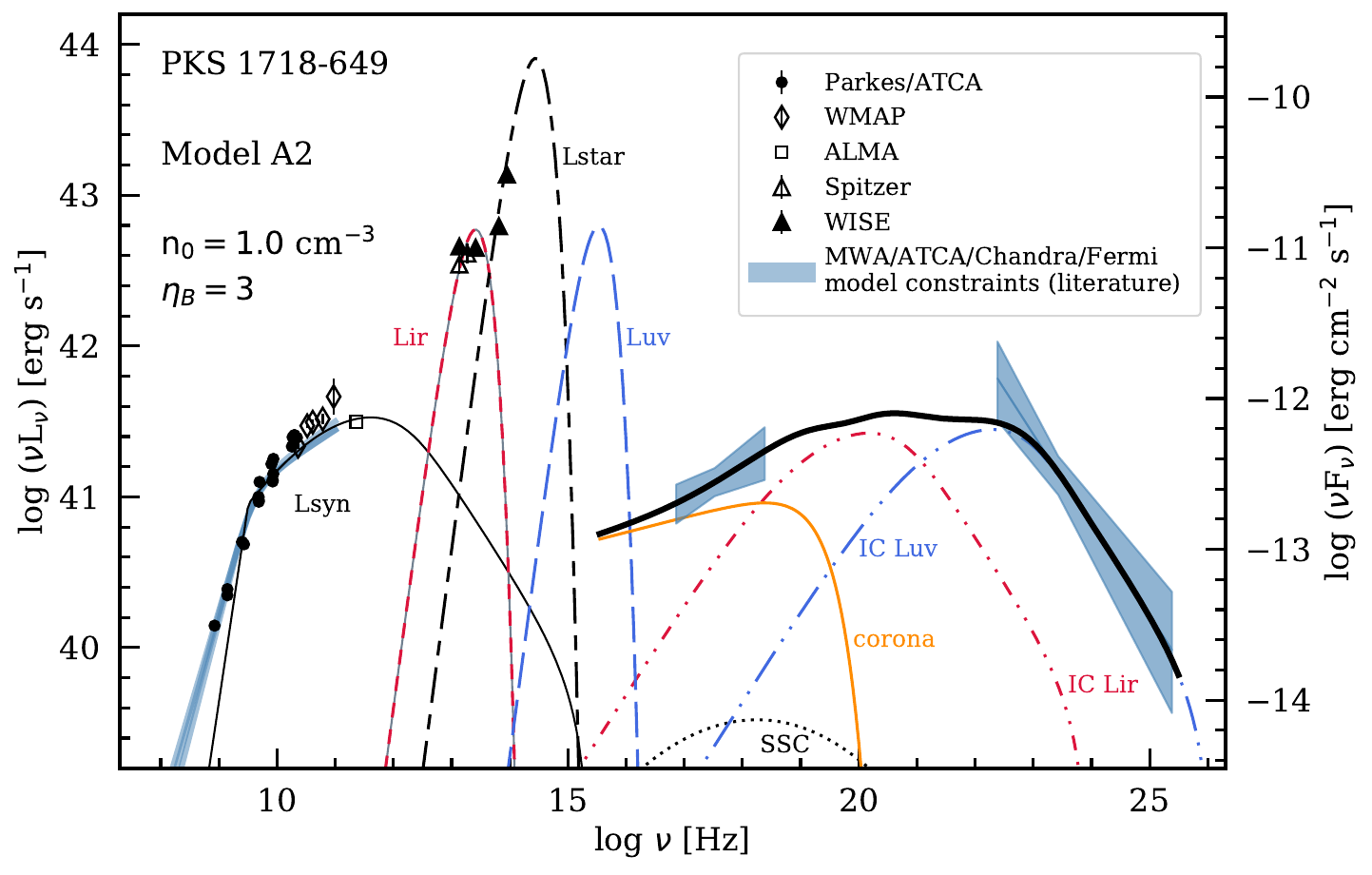}\\
\includegraphics[width=8.8cm]{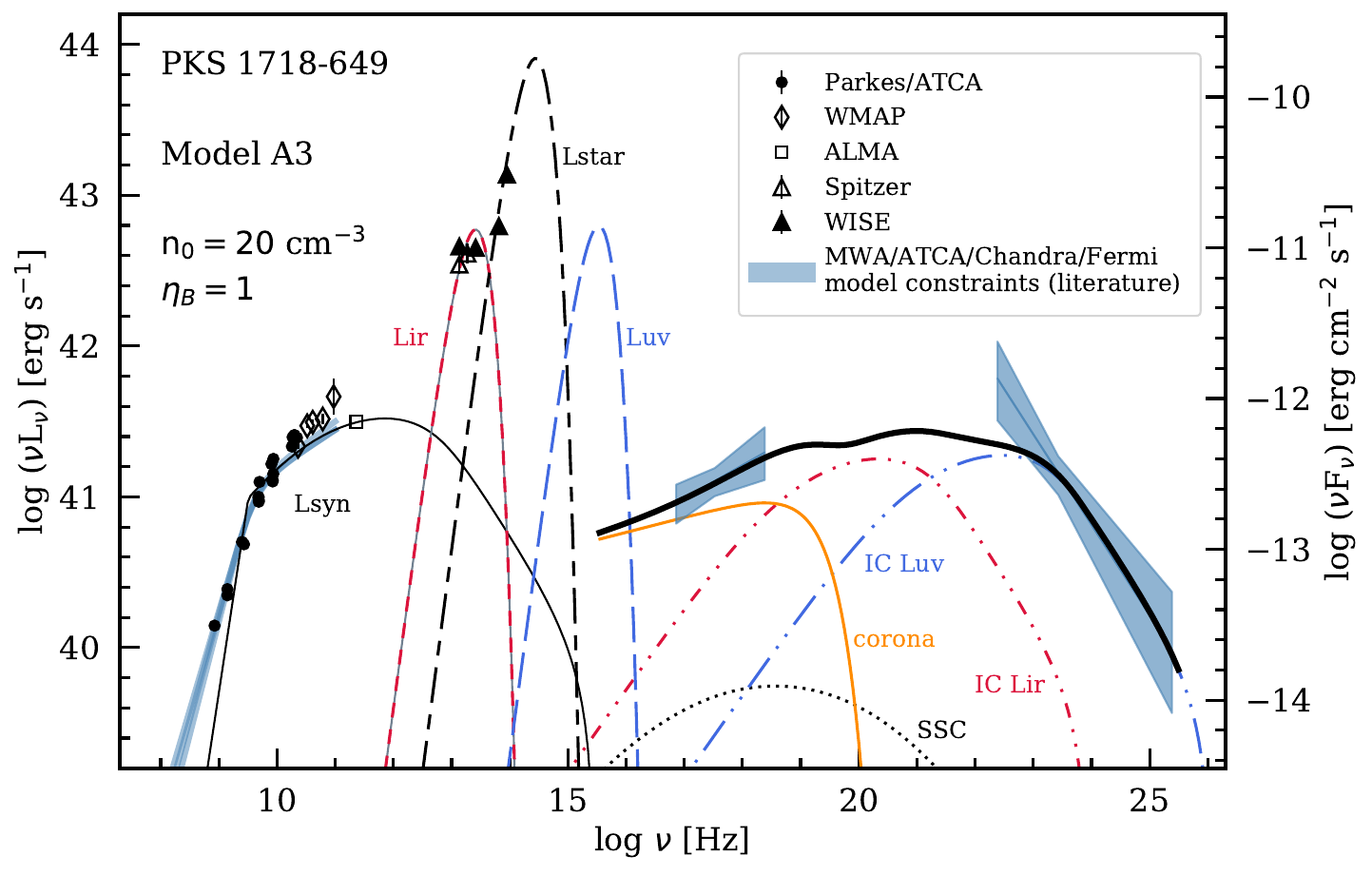}
\includegraphics[width=8.8cm]{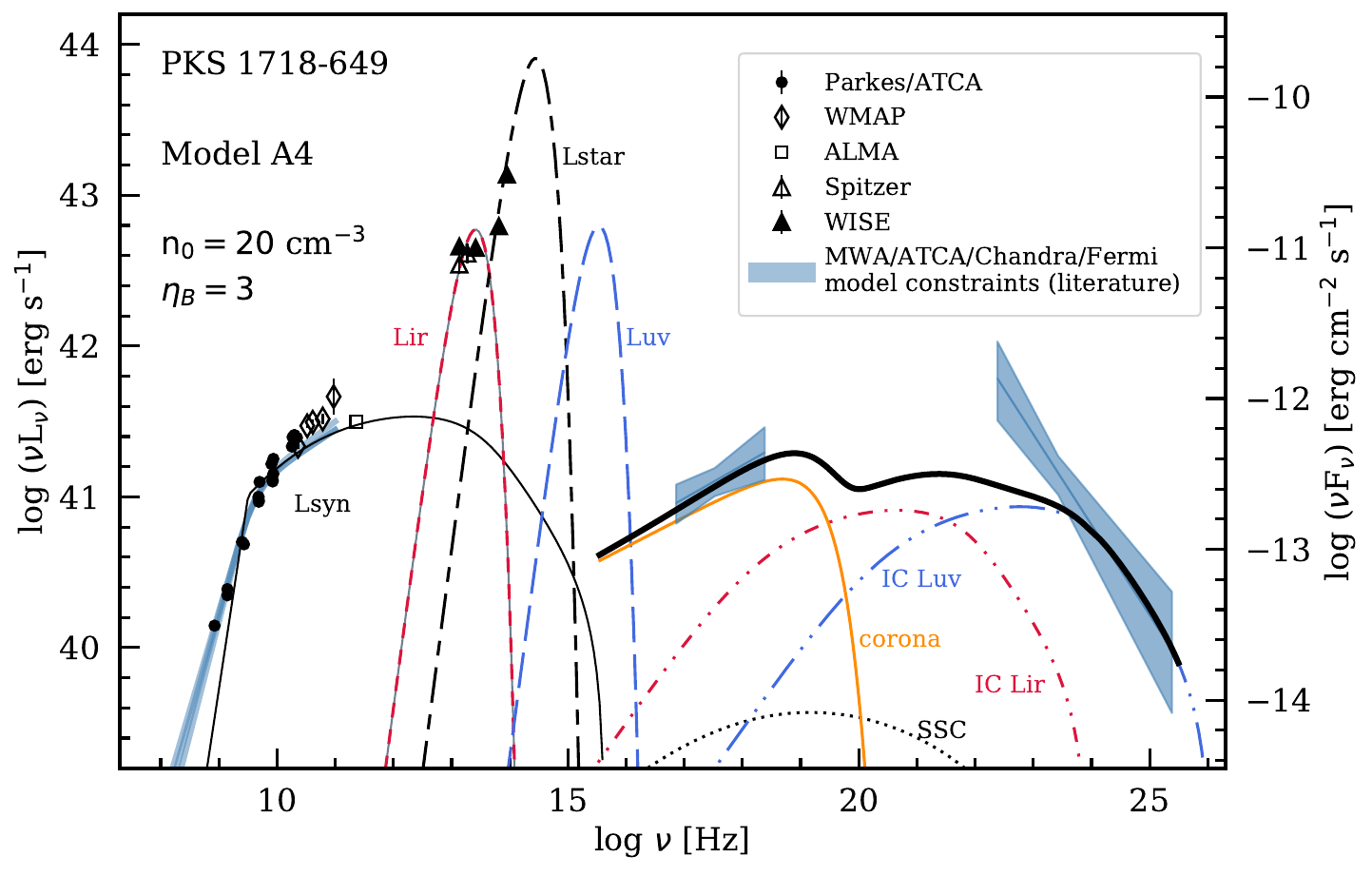}
\caption{\label{fig:sed_appendix} The same as Figure~\ref{fig:sed} but for Model A1 with $n_0 = 3.0$\,cm$^{-3}$ and $\eta_B = 0.3$; Model A2 with $n_0 = 1.0$\,cm$^{-3}$ and $\eta_B = 3$; Model A3 with $n_0 = 20$\,cm$^{-3}$ and $\eta_B = 1$; and Model A4 with $n_0 = 20$\,cm$^{-3}$ and $\eta_B = 3$. See Tables~\ref{tab:sed_fixed} and \ref{tab:sed_appendix} for model parameters.}
\end{center}
\end{figure*}

Supplementary model solutions are presented in Figure~\ref{fig:sed_appendix},
and supplementary model parameters are listed in Table~\ref{tab:sed_appendix} (see
the discussion in Section~\ref{sec:disc}).


\begin{table*}[h]
{\scriptsize
\noindent
\caption[]{\label{tab:sed_appendix} Variable parameters of the supplementary SED models}
\begin{center}
\begin{tabular}{llllllll}
\hline\hline
 Description   & Symbol    & Model A1 & Model A2 & Model A3 & Model A4 & Unit     & Refs.$^a$ \\
\hline
\multicolumn{8}{c}{Injected electron population $Q(\gamma)$ and jet kinetic power} \\ \hline
Low-energy slope              & $\rm s_1$          & 2.1 & 1.8 & 1.9 & 1.9 & & ($\star$)$^b$ \\
High-energy slope             & $\rm s_2$          & 3.3 & 3.2 & 3.2 & 3.1 & & ($\star$) \\
Lorentz factor (min)          & $\gamma_{\rm min}$ & 1   & 1 & 1 & 1  & & (1) \\
Lorentz factor (break)        & $\gamma_{\rm b}$   & 3   & 3.5 & 6 & 12 & $\rm m_p/m_e$ & ($\star$) \\
Lorentz factor (max)          & $\gamma_{\rm max}$ & 100 & 100 & 100 & 100 & $\rm m_p/m_e$ & (2) \\ 
Jet kinetic power             & $L_{\rm j}$        & 5.5 & 1.2 & 1.2 & 0.7 & 10$^{42}$\,erg\,s$^{-1}$ & ($\star$) \\ \hline
\multicolumn{8}{c}{Environment} \\ \hline
ISM density                      & $n_0$           & 3   & 1 & 20 & 20 & cm$^{-3}$                & ($\star$) \\
Electrons                        & $\eta_E$        & 3   & 3 & 3 & 3 &  & (1)$^c$ \\
Magnetic field                   & $\eta_B$        & 0.3 & 3 & 1 & 3 &  & (1, $\star$)$^c$ \\
\hline
\multicolumn{8}{c}{Additional power law X-ray emission} \\ \hline
Photon index                     & $\Gamma$            & not required & 1.9 & 1.9 & 1.8 &  & ($\star$) \\
Luminosity (2-10 keV)            & $L_{\rm 2-10\,keV}$ & n/a & 1.3 & 1.3 & 1.7 & 10$^{41}$\,erg\,s$^{-1}$  & ($\star$) \\
\hline\hline
\end{tabular}
\end{center}
\smallskip Notes:\\
$^a$ References: ($\star$) This work; (1) Stawarz et al. (2008); (2) Ostorero et al. (2010).
$^b$ The radio spectral slope $\alpha\simeq 0.7$ (Tingay et al. 2015) suggests
${\rm s_1} = 2\alpha +1 \simeq 2.4$ for the {\it evolved} electron distribution.
$^c$ Electron and magnetic field energy densities are parametrized as $U_E = \eta_E p$ and
$U_B = \eta_B p$, where $p$ stands for the expanding cocoon's internal pressure.
}
\end{table*}

\end{document}